%% file: u1higgs.tex
\documentclass[12pt,a4paper]{article}
\usepackage{latexsym,graphicx}
\newcommand{\be}{\begin{equation}}
\newcommand{\ee}{\end{equation}}

\newcommand{\bea}{\begin{eqnarray}}
\newcommand{\eea}{\end{eqnarray}}

\def\x{{\bf x}}

\begin{document}
\title{Quasi-Particle Excitations\\
of the Higgs Vacuum at Finite Energy Density
} 

\author{{D. Sexty$^{1}$ and A. Patk{\'o}s$^{2}$}\\
{Department of Atomic Physics}\\
{E{\"o}tv{\"o}s University, Budapest, Hungary}\\
}
\vfill

\footnotetext[1]{denes@achilles.elte.hu}
\footnotetext[2]{patkos@ludens.elte.hu}
\maketitle
\begin{abstract}
\noindent Real time excitations  in the broken symmetry phase
of the classical Abelian Higgs model are investigated numerically
in the unitary gauge. Spectral equations of state of its constituent
quasi-particles are extracted. Characteristic differences between the 
statistical correlations of the
transverse and of the longitudinal vector modes with the Higgs field 
are exposed. A method is proposed for the
reconstruction of the quasi-free composite field coordinates.
\end{abstract}

\section{Introduction}
The rearrangement of the multiplet structure of elementary fields
in Gauge\-+Higgs models in the process of symmetry breaking is one of the most
fundamental concepts upon which the Standard Model of elementary
particle interactions is built \cite{higgs64,higgs66}. The
transmutation of the pseudo-Goldstone field into the longitudinal
polarisation state of massive vector bosons {\it at the minimum of the
  effective potential} has been successfully used in perturbative
calculations. The absence of light (massless) fields from the spectra
of the electroweak sector of the Standard Model
was also tested non-perturbatively first at relatively strong coupling
 \cite{evertz86}, later close to the continuum \cite{csikor96,csikor99}.
Real time investigations of classical gauge field dynamics
mostly concentrated on topological aspects of the process of
baryogenesis, e.g. measuring Chern-Simons densities
\cite{ambjorn85,moore99}, and on the emergence and the
evolution of gauged strings
\cite{rajantie01,shellard01,bellido04}.
 
The idea of far-from-equilibrium baryogenesis
\cite{bellido99,copeland01,copeland02,kuzmin03,smit03} led to increasingly 
detailed studies of the real time mechanism of the reheating of the
Universe at the end of inflation. Features of the parametric
resonance and tachyonic instability were first studied in
simplified inflaton+Higgs systems
\cite{linde94,kofman97,kofman01,borsanyi03},
realising the hybrid inflation scenario. In these systems the 
breaking of the global symmetry 
leads to the generation of Goldstone fields and related
global topological excitations (e.g. strings). Recent numerical
studies of Higgs systems started the detailed
quantitative exploration of the real time excitation process of the
 Higgs and massive vector particles \cite{smit02,smit03,bellido04}. 

J. Smit {\it et al.} \cite{smit02,smit03} developed methods for the
identification of the different physical degrees of freedom in the
process of excitation in Coulomb
and unitary gauges with help of measuring the corresponding
 dispersion relations. 
They measured the time variation of the corresponding number
densities during tachyonic (spinodal) instability. In the unitary gauge
they find significant differences between
the early values taken by the number densities of the transversal and 
longitudinal degrees
of freedom.  The differences diminish when the
system approaches equilibrium. 

Our final goal in the real time investigation of the Abelian Gauge\-+Higgs
system is to measure the evolution of the equations of state of its
physical components during the reheating period. To achieve this we
 apply methods
developed in our earlier investigation of pure scalar (inflaton+Higgs) systems
\cite{borsanyi02,borsanyi03}. In order to understand the nature of the
final state to which the system eventually relaxes in this paper a careful 
 numerical study of the real time fluctuations in equilibrium systems
with low energy densities is presented. 
We extract the equilibrium equations of state of the different
quasi-particle species and an attempt will be described to establish
the corresponding effective field-coordinates. Such
dynamical variables are present in rather strongly interacting
systems of condensed matter physics, even when the original particles
might decay into each other. 
 Of course, this investigation could be performed also in the Euclidean
formulation of equilibrium field theories on lattice. 
For the clarification of specific
issues already at this stage the possibility of real time studies of 
the non-equilibrium time evolution was quite useful (see below).

We solved numerically the field equations of the Abelian Higgs model,
discretized on a spatial lattice in the axial $(A_0=0)$ gauge. The
initial conditions were chosen with very low energy density in order
to end up in the broken symmetry phase.  In the present
investigation mainly the couplings $e^2=1, \lambda=6$ were used.
The influence of the variation of the couplings on
our findings was systematically explored in a broad range.

As a simple (global) signature of reaching near thermal equilibrium
we have chosen the equality of the energy density stored in the scalar
field (defined as part of the Hamiltonian density independent of the
gauge fields) with one third of the energy density in gauge modes. 
It took surprisingly long time to reach stationary equality: 
 only after 
$t \approx 4 \times 10^5 |\textrm{m}|^{-1} $  did fluctuations in the
 difference of the energy densities settle at the one percent level.
Then characteristic thermodynamical quantities were analysed for an
extended time interval, sufficiently long to explore the thermal
features. It is remarkable that the equations of state, to be discussed
 below, take their equilibrium form
considerably before the true equilibrium is reached. This experience
supports the findings of \cite{borsanyi03,berges04}.

The equilibrium configurations were transformed to the
unitary gauge and decomposed into a scalar (Higgs) field, plus
transversal and longitudinal gauge fields. 
The full energy density and pressure of the system can be split up in
a rather natural way into pieces associated with these fields. In
section 2 we shall present the method of constructing spectral decompositions
for these quantities without making any assumption about the nature of
the statistically independent field variables. Still, we can convincingly
argue that the quasi-particle interpretation of the thermodynamics
works and the extracted mass degeneracy agrees with the result of the weak
coupling analysis.

In section 3,  evidence will be shown that the transverse vector modes behave
as independent Gaussian statistical variables. On the other hand, the
longitudinal vector mode, defined with the usual projection violates
the expected Gaussian relation between the quadratic and quartic
moments and does not average independently from the Higgs mode even
when the average energy per degree of freedom was one thousendth of
the Higgs mass.  The possible dependence of this
characteristic difference on the Higgs vs. vector mass relation (that
is on the couplings $e^2, \lambda$) was carefully explored. Also the
effect arising from the presence of vortex-antivortex pairs was
understood with help of studying the non-equilibrium time evolution of
the system.

The persistent interaction between the longitudinal gauge and the
Higgs field motivates the search for the quasi-particle
''coordinates'' in terms of non-linear combinations of these fields.
Section 4 presents the results of the first step made in this direction. 
 In our search for a Gaussian field related to the longitudinal vector
 mode, and
 fluctuating independently from the other three fields, a new pair of 
canonical variables is proposed:
\be
{\bf {\cal A}}_L^{(1)}(\x,t)=(1+\alpha\rho^2(\x,t)){\bf A}_L(\x,t),
\quad {\bf {\cal P}}_L^{(1)}(\x,t)=\frac{1}{(1+\alpha\rho^2(\x,t))}
{\bf \Pi}_L(\x,t).
\label{nonlin}
\ee 
The value of the coefficient $\alpha$ was determined by the
requirement to have for ${\cal A}_L ({\cal P}_L)$ and $\rho$ independent
statistics. Having done this the new variables are checked for
Gaussianity.  Also an alternative composite field of the form
\be
{\cal A}_L^{(2)}({\bf x},t)=\rho({\bf x},t)^\alpha{\bf A}_L({\bf x},t),
\qquad {\cal P}_L^{(2)}({\bf x},t)=\rho({\bf x},t)^{-\alpha}
{\bf\Pi}_L({\bf x},t)
\label{nonlin2}
\ee
will be tested with satisfactory results. The change of variables
influences only very mildly the Higgs field itself. This is rather
fortunate since this variable, as well as the transverse gauge degrees of
freedom, consistently obeys free quasi-particle thermodynamics.
 Further possible tests of the existence of a Gaussian
composite field involving the 
longitudinal vector mode are discussed in the Conclusions.
Technical details of the real time numerical investigation are
summarized in an Appendix.

\section{The thermodynamics of
  the Abelian Higgs model}

In the first part of this section we review the continuum expressions of
the relevant thermodynamical quantities of the Abelian Higgs model in
the unitary gauge. A model-independent spectral test is performed 
which confirms that in the time interval where the analysis of the
thermal features is realized the energy is equally partitioned in momentum
space among the different modes. After checking that the potential energy
density and the kinetic energy of the spectral modes are equal in
equilibrium this fact is used to derive simple expressions for the
spectral equations of state determining the thermodynamics of
 the Higgs, the longitudinal and the transverse vector modes. The
 mass degeneracy of the different vector polarisations is demonstrated.
 The corresponding effective masses are close to the tree level expectations. 

The Lagrangian of the Abelian Higgs model is given by the expression
\be
L=-\frac{1}{4}F_{\mu\nu}F^{\mu\nu}+\frac{1}{2}D_\mu\Phi(D^\mu\Phi)^*-V(\Phi),
\ee
where $F_{\mu\nu}=\partial_{[\mu}A_{\nu]},
D_\mu=\partial_\mu+ieA_\mu$.
$V(\Phi)$ is the usual quartic potential of the complex Higgs field
$\Phi$:
\be
V(\Phi)=\frac{1}{2}m^2|\Phi|^2+\frac{\lambda}{24}|\Phi|^4,\qquad m^2<0.
\ee
The result of the standard calculation of the energy-momentum tensor
for the Abelian Higgs model leads in the unitary gauge to the
following decomposition for the energy density:
\bea
\epsilon&=&\epsilon_\rho+\epsilon_T+\epsilon_L,\nonumber\\
\epsilon_\rho&=&\frac{1}{2}\Pi_\rho^2+\frac{1}{2}(\nabla\rho)^2+V(\rho),
\nonumber\\ 
\epsilon_T&=&\frac{1}{2}[{\bf \Pi}_T^2+(\nabla\times{\bf
    A}_T)^2+e^2\rho^2{\bf A}_T^2],\nonumber\\
\epsilon_L&=&\frac{1}{2}[{\bf \Pi}_L^2+e^2\rho^2({\bf A}_L^2+A_0^2)].
\label{energydensities}
\eea
Here the indices $T~(L)$ refer to the fact that the corresponding
terms contain the transverse (longitudinal) component of the vector
field and its canonically conjugated momentum.
The energy density is written in Hamiltonian formulation employing the 
relations:
\be
\Pi_\rho=\dot\rho,\quad {\bf\Pi}_T=\dot{\bf A}_T,\quad {\bf
  \Pi}_L=\dot{\bf A}_L+\nabla A_0.
\ee
Also one has to understand $A_0$ as a dependent variable in view of the
Gauss-condition: 
\be
\nabla{\bf \Pi}_L=e^2\rho^2A_0.
\ee
The space-space diagonal components of the energy-momentum tensor give 
similar expressions for the partial pressures:
\bea
p&=&p_\rho+p_T+p_L,\nonumber\\
p_\rho
&=&\frac{1}{2}\Pi_\rho^2-\frac{1}{6}(\nabla\rho)^2-V(\rho),\nonumber\\
p_T&=&\frac{1}{6}[{\bf \Pi}_T^2+(\nabla\times{\bf
    A}_T)^2-e^2\rho^2{\bf A}_T^2],\nonumber\\
p_L&=&\frac{1}{6}[{\bf \Pi}_L^2-e^2\rho^2{\bf A}_L^2]+
\frac{1}{2}\frac{1}{e^2\rho^2}(\nabla{\bf\Pi}_L)^2.
\label{pressures}
\eea

\begin{figure}
\begin{center}
\includegraphics[width=12cm]{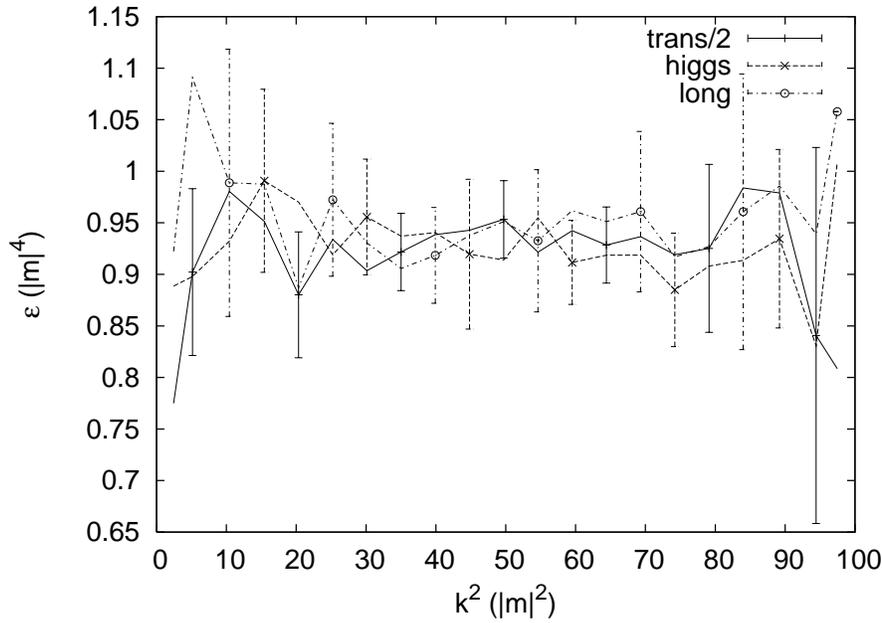} 
\end{center}
\caption{ Energy densities of the Higgs and gauge fields in equilibrium,
corresponding to $\epsilon_{total}= 3.77 |m|^4 $. The data points and 
the error bars
were obtained by binning and forming the mean energy per mode and its 
fluctuation for each bin.
  }
\label{kep:ekvi}
\end{figure}

Spectral energy densities can be introduced for each of the three
contributions defined in Eq.(\ref{energydensities}) 
by considering the square root of each
$\epsilon_i({\bf x},t),~i=\rho, T, L$ and taking  the absolute square
of their Fourier-transform:
\bea
\overline{\epsilon}_i(t)&=&\frac{1}{V}\int d^3x\epsilon_i({\bf
  x},t)=\int \frac{d^3k}{(2\pi)^3}|\epsilon_i^{(1/2)}({\bf
  k},t)|^2,\nonumber\\
\sqrt{\epsilon_i({\bf x},t)}&=&\int\frac{d^3k}{(2\pi)^3}e^{i{\bf
    kx}}\epsilon^{(1/2)}_i({\bf k},t).
\eea
There is some ambiguity in factoring the densities into two \textit
{identical} powers.
This choice is the most natural for free
quasi-particle interpretation of the thermodynamics, since those
expressions are quadratic in the effective field-coordinates.
In the broken symmetry phase at equilibrium one expects
$|\epsilon^{(1/2)}_i({\bf k},t)|^2$ to fluctuate around $\bf
k$-independent values obeying the relation:
\be
\overline{|\epsilon^{(1/2)}_L({\bf k})|^2}:
\overline{|\epsilon^{(1/2)}_T({\bf k})|^2}:
\overline{|\epsilon^{(1/2)}_\rho({\bf k})|^2}=1:2:1.
\ee
The overline means an average over the different modes with the same
length $k$, and over a certain time-interval.
This was found to be satisfied with 10\% relative fluctuation in the
relevant time interval as illustrated in Fig. \ref{kep:ekvi}. When
below we refer
to a sort of ''temperature'' we have in mind the average energy
density in the $\bf k$-space. If a free quasi-particle model works,
then this temperature is twice the average kinetic energy per mode.

Similar construction can be performed also for the partial pressure of
the $i$-th constituent, providing a model-independent definition for
its spectral decomposition: 
\be \overline{p_i}({\bf k})\equiv
\overline{|p^{(1/2)}_i({\bf k})|^2}.  
\ee 
The central step in the analysis is the
investigation of the \textit{spectral equation of state} defined as 
\be
w_i({\bf k})\equiv \frac{\overline{p_i}({\bf
    k})}{\overline{\epsilon_i}({\bf k})}.  
\label{specteqstate}
\ee 
This quantity can be measured for each configuration.  In the equilibrium
the spectral densities of the kinetic and of the potential energy
are equal on average. Using
this feature as a hypothesis for the Fourier-modes a very simple
functional form is predicted in case, for instance, of the Higgs-mode: 
\be
w_\rho({\bf k})=\frac{1}{3}\frac{{\bf k}^2\overline{|\rho({\bf k})|^2}}{{\bf
k}^2\overline{|\rho({\bf k})|^2}+2\overline{V_\rho({\bf k})}}. 
\ee 
If the Higgs-field is a
''pure quasi-particle coordinate'', then it performs small amplitude
oscillations near its equilibrium value $\rho_0$: 
\be
V(\rho)=\frac{1}{2}M_{\rho,eff}^2(\rho-\rho_0)^2,\qquad V_\rho({\bf
k})=\frac{1}{2}M_{\rho,eff}^2|\rho({\bf k})|^2,\quad k\neq 0.  
\ee
This leads to a one-parameter ''free-field'' 
form of its spectral equation of state:
\be 
w_\rho({\bf k})=\frac{1}{3}\frac{|{\bf k}|^2}{|{\bf
k}|^2+M_{\rho,eff}^2}.  
\label{eqstatrho1}
\ee 
Similar line of thought can be followed
also in the spectral analysis of the transversal gauge energy density and
pressure. The result is the following: 
\be w_T=\frac{1}{3}\frac{{\bf k}^2}{{\bf k}^2+M_{T,eff}^2},\qquad
M_{T,eff}^2=\frac{\overline{|(e\rho{\bf A}_T)({\bf k})|^2}}
{\overline{|{\bf A}_T({\bf k})|^2}}.
\label{eqstattrans1}
\ee
If one finds a good description of the measured spectral equation of
state by this formula with a ${\bf k}$-independent mass, this
provides evidence for the free quasi-particle interpretation of the
transverse part of the thermodynamics.

For the longitudinal polarisation a very similar formula can be
obtained when the ratio $w_L({\bf k})$ is defined by the scaled energy
density $e^2\rho^2\epsilon_L$ and pressure $e^2\rho^2p_L$ as
\be
w_L({\bf k})\equiv\frac{\overline{|(e\rho \sqrt{p_L})({\bf k})|^2}}
{\overline{|(e\rho \sqrt{\epsilon_L})({\bf k})|^2}}.
\label{eqstatlong1}
\ee
It leads to the above quasi-particle form if the mode-by-mode
equality of the scaled kinetic and potential energies is obeyed in the form
\be
 {\bf k}^2\overline{|{\Pi_L}({\bf k})|^2}+\overline{|(e\rho\Pi_L)({\bf
     k})|^2}=
\overline{|(e^2\rho^2{\bf A}_L)({\bf k})|^2},
\label{equilong}
\ee
with the following effective mass formula:
\be
M_{L,eff}^2=\frac{\overline{|(e\rho\Pi_L)({\bf k})|^2}}
{\overline{|\Pi_L({\bf k})|^2}}.
\ee
   
The first task was therefore to check numerically if the average
equality of the kinetic and potential mode energies is
fullfilled (for the longitudinal case, Eq.(\ref{equilong}), see 
Fig. \ref{kep:ekvipart}). 
Next, the measured $w_i({\bf k})$ functions were fitted with
$k$-independent effective masses. Finally their values were compared to
the tree-level estimates:
\be
M^2_{L,eff}=M^2_{T,eff}=e^2\overline{\rho^2},\quad
M_{\rho,eff}^2=\frac{\lambda}{3}\overline{\rho^2}.
\label{Lagr-mass}
\ee

\begin{figure}
\begin{center}
\includegraphics[width=8.5cm]{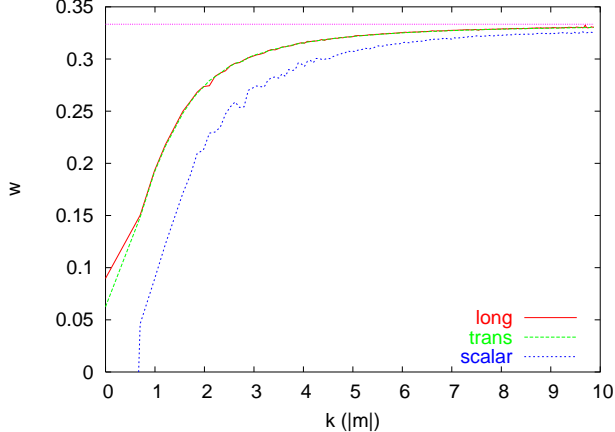} 
\end{center}
\caption{Spectral equations of state computed from
  Eqs. (\ref{specteqstate})  and  (\ref{eqstatlong1})
as applied to the three constituting terms
  of the energy density and pressure.} 
\label{kep:wekvi}
\end{figure}

Fig.\ref{kep:wekvi} displays the measured 
equations of state of the three types of fields. It is obvious that
the longitudinal and the transverse gauge fields show degenerate
behaviour despite of the rather different formal procedure of the
determination of $w_L$ and $w_T$. The fitted squared 
masses appear in Table \ref{tabl1}.

\begin{table}[hbt!]
\begin{center}
\begin{tabular}{|c|c|c|c|c|c|c|}
\hline
 & & \multicolumn{2}{c|}{Lagrangian mass squared} & 
 \multicolumn{3}{c|}{equipartition mass squared} \\ 
\hline
 $ T_{kinetic}(|m|) $  & $\epsilon_{total} (|m|^4)  
$ & scalar & trans \& long & scalar & trans & long \\
\hline
\input masstabl2.tab
\hline
\end{tabular}
\end{center}
\caption{The comparison of the averaged mass terms of the Lagrangian
  (e.g. $M_i^2\varphi_i^2, ~~ \varphi_i=\rho,{\bf A}_T,{\bf A}_L$) with 
the thermodynamically fitted effective squared masses $M_{i,eff}^2$.}
\label{tabl1}
\end{table}

It is remarkable that for the Higgs field the
 measured values of the squared mass are 10-15\% higher than expected
 on the basis of (\ref{Lagr-mass}). This is the effect of the
 ultraviolet fluctuations which modify at one loop its mass 
\cite{copeland01} the following way:
\be
M^2_{H,lattice}=m^2+\frac{\lambda}{2}\overline{\rho^2}+\left(3e^2+
\frac{\lambda}{2}\right)\frac{0.226}{a^2}.
\ee
Indeed, the variation of the deviation from the tree level squared
 mass at very low energy densities
was found to be inversely proportional to $a^2$. The
 non-perturbatively fitted coefficient is much smaller, than the 1-loop
 estimate appearing above. The convergence of the measured Higgs mass
to a well-defined value for $T\approx 0$ makes it possible to define
 the renormalised 
Higgs mass. If one chooses it to coincide with the curvature of the 
classical potential at its minimum (e.g. (\ref{Lagr-mass})),
then with the same subtraction one can define also the temperature
dependent mass. In this way a very good agreement of the renormalised
 squared masses with the predicted temperature dependence was found.

The effective squared masses of the vector modes were found very close to
$M^2_V=\overline{e^2\rho^2}$ without observable
lattice spacing dependence.

\section{The statistics of the elementary fields}

\begin{figure}
\begin{center}
\includegraphics[width=8.5cm]{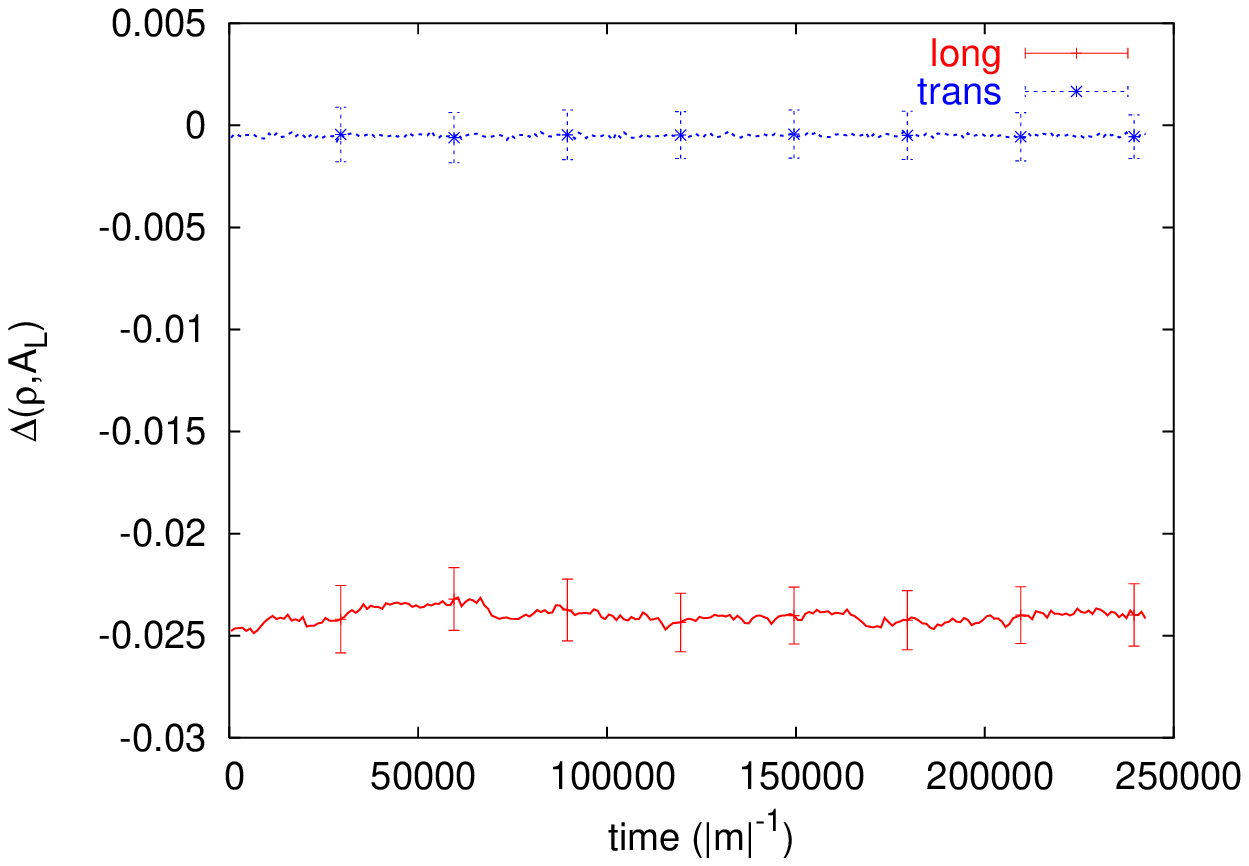} 
\includegraphics[width=8.5cm]{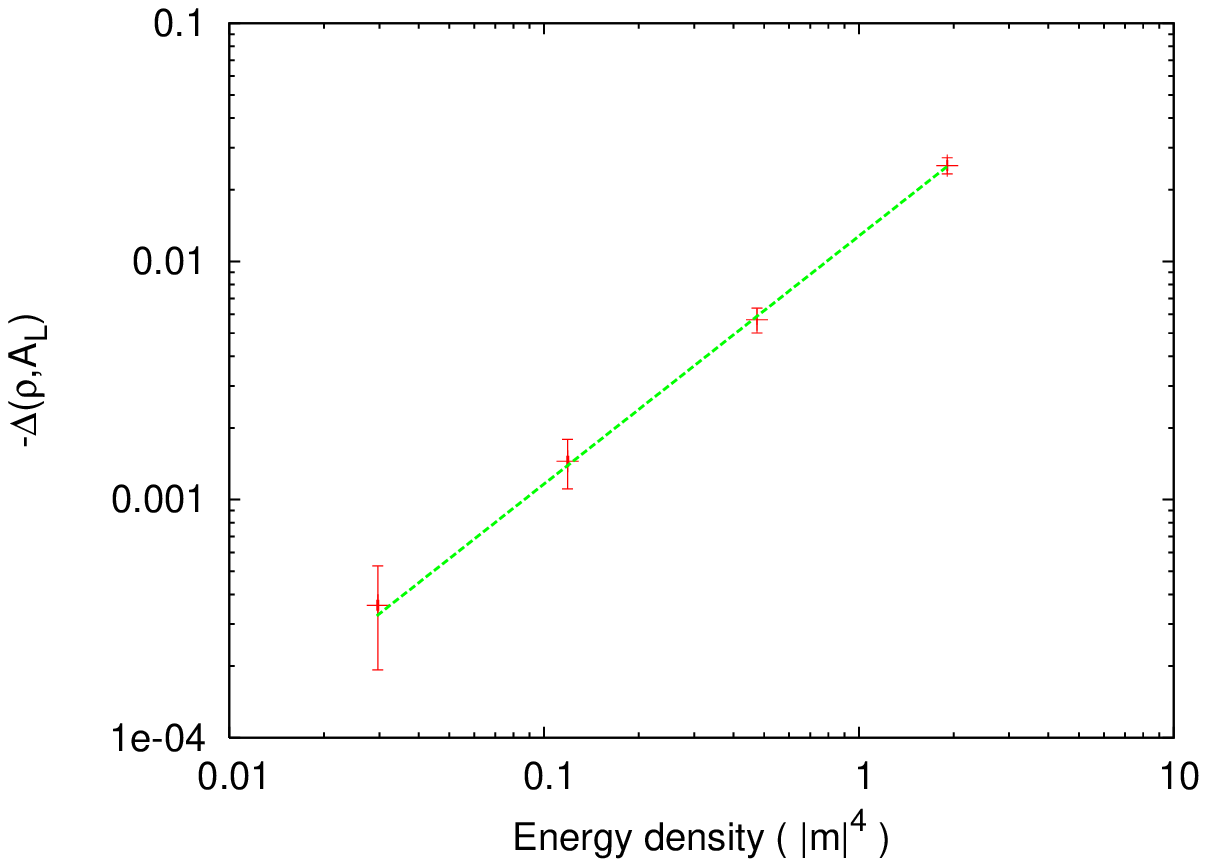}
\end{center}
\caption{Correlation between the scalar field and the longitudinal and
  transversal gauge fields.
On the left the binned averages of (\ref{nonfactlong}) are
  displayed as a function of time (measured in units $|m|^{-1}$). 
The error bars show the actual  amplitude of its fluctuations. 
 The average values of the correlation coefficients
  are shown in the right hand figure as a function of the energy
  density (measured in units $|m|^4$). 
The variation is compatible with the vanishing of the
  correlation for vanishing energy density
 (its slope is: $ 1.04 \pm 0.04  $). }
\label{kep:elso}
\end{figure} 

In this section we shall analyse the data in further details
focusing on the statistical correlations
of the field variables $\rho, {\bf A}_L, {\bf A}_T$. Our aim is to
find the field-coordinates of the quasi-particles, emerging from
the thermodynamical analysis of the previous section. There are two
aspects of this search:
 
\begin{itemize}
\item{} Can one consider the above variables mutually independent in
  statistical sense?
\item{} Do they follow Gaussian statistics expected for small
 amplitude, nearly free quasi-particle oscillations?
\end{itemize}
As a signature of independence of two field variables we shall consider
the factorisability of the spatial average of their quadratic products
by testing if, for instance, 
\be
\Delta[{\bf A}_T,\rho]\equiv
\frac{\overline{\rho^2({\bf x},t){\bf A}_T^2({\bf x},t)}-
\overline{\rho^2({\bf x},t)}~\overline{{\bf A}_T^2({\bf x},t)}}
{\overline{\rho^2({\bf x},t){\bf A}_T^2({\bf x},t)}} =   0
\label{nonfactlong}
\ee
is fullfilled (analoguous test is performed for $\Delta[{\bf A}_L,\rho]$).
For variables obeying Gaussian statistics the equality
\be
\Gamma[{\bf A}_T]\equiv\frac
{\overline{\left({\bf A}_T^2({\bf x},t)\right)^2}-3\left(
\overline{{\bf A}_T^2({\bf x},t)}\right)^2}
{\overline{\left({\bf A}_T^2({\bf x},t)\right)^2}}= 0,\nonumber\\
\label{nonGauss}
\ee
is obeyed.

 Clear quantitative evidence appears on the left hand 
side of Fig.\ref{kep:elso} for the stationary (time-independent)
independence of ${\bf A}_T$ and  $\rho$.
On the other hand ${\bf A}_L$ and $\rho$ show non-zero
correlation, which appears also constant in time.
 The test of Gaussianity shows that from very early times
both $\Gamma[{\bf A}_T]$ and $\Gamma[\rho]$ vanish, while $\Gamma[{\bf
    A}_L]$ violates the criterium concerning the Gaussian nature of
its statistics.  

 On the right hand side of Fig. \ref{kep:elso} we give the 
dependence of $\Delta[{\bf A}_L,\rho]$ on the average energy density of the
system. (The corresponding temperatures,  understood in the sense
described in section 2, are listed in Table 1).
 The effect increases linearly with the energy density. 
Analogous behaviour is observed for $\Gamma[{\bf \Pi}_L]$ and
$\Delta[{\bf \Pi}_L,\rho]$, where ${\bf \Pi}_L$ is the canonical momentum
field conjugate to ${\bf A}_L$. The correlation coefficient $\Delta ({\bf
  A}_L,\rho)$ was computed also on lattices of the same physical size
but smaller lattice spacing providing essentially the same result. 
The lack of the $\rho - {\bf A}_L$ factorisation was demonstrated
on a number of functions $f(\rho)$ replacing $\rho^2$, among them also 
$(\rho-\overline{\rho})^2$.

\begin{figure}
\begin{center}
\includegraphics[width=8.5cm]{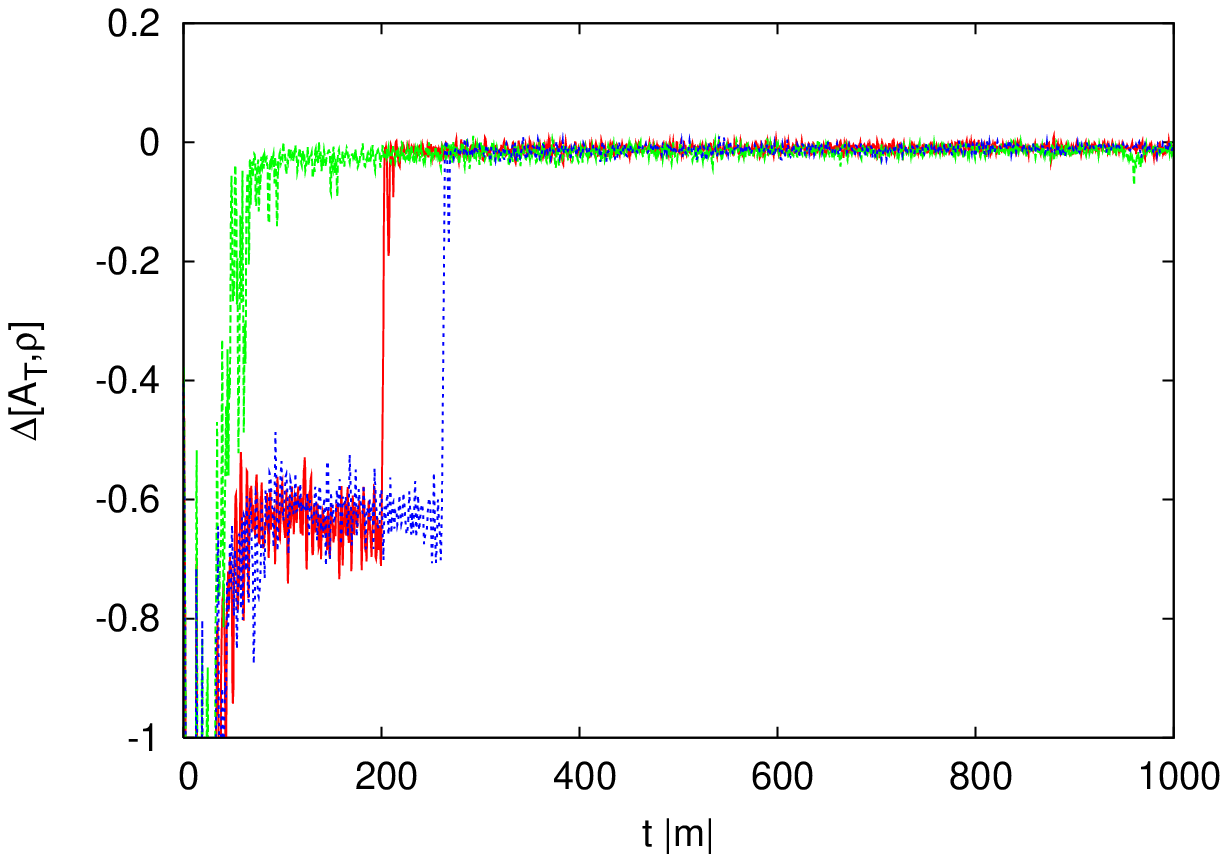} 
\includegraphics[width=8.5cm]{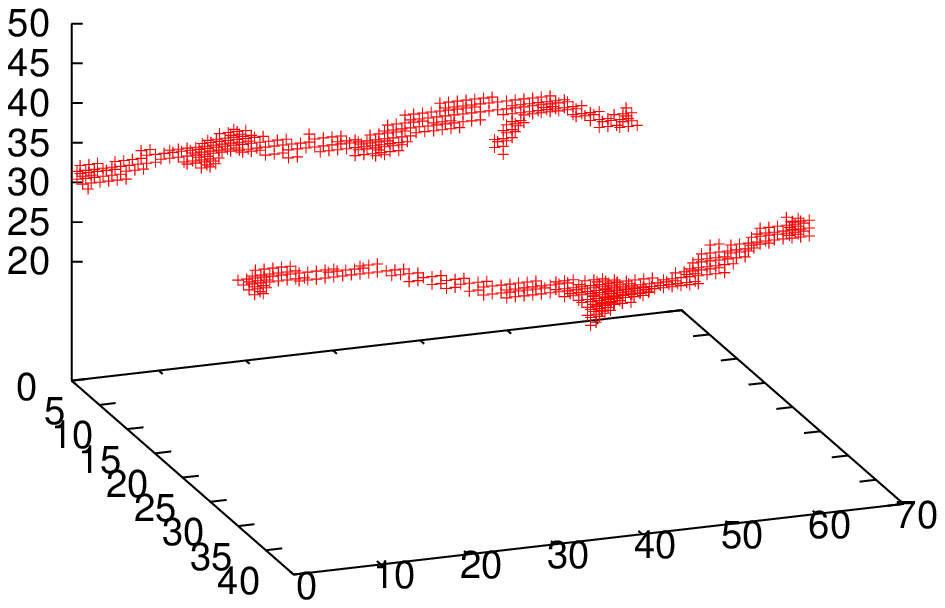}
\end{center}
\caption{Left: Three examples of the non-equilibrium evolution of
  $\Delta[{\bf A}_T,\rho]$. In two time-histories
 stationary nonzero values of $\approx -0.6$
  occured for a restricted time interval. It signals the
  presence of a vortex-antivortex pair and can be easily distinguished
  from the fast and smooth relaxation to zero, when no vortices are
  generated. Right: The 
vortex-antivortex pair  corresponding to non-zero $\Delta$, 
is found by looking for the points where $\rho < 0.3$}
\label{kep:transverse-korr}
\end{figure}

In systems with large values of the couplings this strong correlation 
could be natural to expect. In this respect it is more surprising that the
    transversal gauge field behaves as a perfect independent variable.
In non-equilibrium time evolution, when in addition to the small
amplitude white noise initial conditions applied to the ${\bf k}\neq 0$ modes,
the average Higgs field  (${\bf k}=0$) starts
    from the unstable maximum at $\overline{\rho}=0$ we have observed
    frequently the generation of rather large nonzero values for   
$\Delta[{\bf A}_T,\rho]$. Some spectacular examples are
    presented in Fig.\ref{kep:transverse-korr}. One observes a rather
    abrupt vanishing of these values after certain time elapses. It turns
    out that in case of the transverse fields nonzero quasi-stationary
    values of $\Delta[{\bf A}_T,\rho]$ 
are very sensitive indicators of the presence
    of Nielsen-Olesen vortex-antivortex pairs \cite{olesen73}. A representative
    vortex-antivortex pair is displayed in the right hand figure of
    Fig.\ref{kep:transverse-korr}. The drop of the absolute value of
    $\Delta[{\bf A}_T,\rho]$ occurs when the pair annihilates. 
    Similar, but less pronounced drop is observed in the correlation
of the Higgs field and the magnetic plaquette variable, which is strictly
gauge invariant even for finite lattice spacing.
    It is
    clear that in the vortex solution presented in \cite{olesen73} 
the transversal vector potential and the real Higgs field vary in a 
correlated way. The detailed statistics of the vortex production
    during  spinodal instability will be discussed in a separate
    publication. Here we conclude as a byproduct 
from this analysis, that the vanishing of $\Delta[{\bf A}_T,\rho]$ in
equilibrium not only hints to the independence of the ${\bf
  A}_T$-statistics, but also represents evidence that no vortex
    lines are present in our equilibrium configurations.

We have investigated also the dependence of $\Delta[{\bf
    A}_L,\rho]$ and $\Delta[{\bf A}_T,\rho]$
on the Higgs-gauge mass ratio, that is on
$\lambda/e^2$. The value of $\lambda$ was varied in the range
$(0.75,75)$ keeping the energy density fixed and $e^2=1$. $\Delta[{\bf
    A}_L,\rho]$ linearly increases from $\sim 0.001$ up to $\sim 0.1$.
Specifically, no effect on this linear variation can be observed when
the decay channel of the Higgs into two ''photons'' opens around
$\lambda/e^2\sim 12$. The time average of $\Delta[{\bf A}_T,\rho]$ 
 stays zero, only the size of its fluctuations increases.
We have checked that the correlative effects 
 do not change for $e^2=0.01$.

In concluding this section we elaborate on the consequences of the
statistical independence of ${\bf A}_T$ and $\rho$ on the
thermodynamics. The Gaussian nature of the Fourier modes of ${\bf
  A}_T$ is expressed by the following formula:
\be
\overline{{\bf A}_T({\bf k}){\bf A}^*_T({\bf k}')}=\overline{|{\bf
    A}_T({\bf k})|^2}\delta_{{\bf k},{\bf k}'}.
\label{at-statistics}
\ee
Let us apply it to the statistical expectation of a weighted space-average
of ${\bf A}_T({\bf x})$. 
One can make the observation that both $\epsilon_T$ and $p_T$ are
quadratic expressions of ${\bf A}_T$, multiplied by some function of
$\rho$. First one substitutes for ${\bf A}_T$ its
Fourier expansion. When the independence of the ${\bf A}_T$ statistics
from $\rho$ is used, the average can be written in a factorized
way. Finally, (\ref{at-statistics}) is used and provides a simple
general formula:
\bea
\overline{\frac{1}{V}\int d^3xf[\rho({\bf x},t)]{\bf A}_T^2({\bf
      x},t)}&=&
\overline
{\frac{1}{V}\int d^3xf[\rho({\bf x},t)]\sum_{{\bf k},{\bf k}'}e^{i({\bf
    k}-{\bf k}'){\bf x}}~}\times\overline{{\bf A}_T({\bf k},t){\bf A}_T^*({\bf
    k}',t)}\nonumber\\ 
&=&\overline{\langle f[\rho({\bf x},t)]\rangle_V}\sum_{\bf k}\overline{|{\bf
      A}_T({\bf k},t)|^2},
\eea
where the bracket with lower index $V$ was introduced to denote
 the spatial averaging.
This representation allows for the following identification of the
corresponding spectral density, with no ambiguity left in its definition:
\be
\overline{\langle f[\rho]{\bf A}^2\rangle_V}({\bf k})\equiv 
\overline{\langle f[\rho({\bf
      x},t)]\rangle_V}~\overline{|{\bf A}({\bf k},t)|^2}.
\ee
Direct consequences of this are the following expressions for
$\overline{\epsilon}_T, \overline{p}_T$, where factorized averaging
over ${\bf A}_T$ and $\rho$ appears:
\bea
\overline{\epsilon_T}({\bf
  k})&=&\frac{1}{2}\left[\overline{|{\bf\Pi}_T({\bf k})|^2}+({\bf
    k}^2+\overline{\langle e^2\rho^2\rangle_V} )
\overline{|{\bf A}_T({\bf k})|^2}\right],
\nonumber\\
\overline{p_T}({\bf k})&=&\frac{1}{6}\left
[\overline{|{\bf\Pi}_T({\bf k})|^2}+({\bf
  k}^2-\overline{\langle e^2\rho^2\rangle_V}) 
\overline{|{\bf A}_T({\bf k})|^2}\right].
\eea
They are equivalently given through the expression of the dispersion relation
and the equation of state of the transversal quasi-particles: 
\be
\omega_T^2({\bf k})=\frac{\overline{|{\bf\Pi}_T({\bf k})|^2}}
{\overline{|{\bf A}_T({\bf k})|^2}}={\bf
  k}^2+\overline{\langle e^2\rho^2\rangle_V},\qquad
w_T({\bf k})\equiv\frac{\overline{p}_T({\bf k})}
{\overline{\epsilon}_T({\bf k})}=\frac{1}{3}\frac{{\bf k}^2}{{\bf
    k}^2+\overline{\langle e^2\rho^2\rangle_V}}.
\label{eqstattrans}
\ee
(The dispersion relation emerges from the average 
kinetic-potential energy equality for each mode).
These expressions reproduce perfectly the original transversal
spectral densities and the EoS extracted in the previous section.

Similarly good representation is obtained for the thermodynamics of
the Higgs-sector when the Gaussianity of $\rho$ is imposed on the
averaging. In case of ${\bf A}_L$ a very bad quality representation of
the spectral equation of state is possible if independence and
Gaussianity are forced upon the longitudinal part of the vector
potential. The quality of the description can be characterized by the
violation of the kinetic-potential energy equality (\ref{equilong}),
which is displayed in Fig.\ref{kep:ekvipart}.

Since the characteristics of the longitudinal equation of state
clearly possess a quasi-particle character, as was seen in section 2,
 our conclusion is that
${\bf A}_L$ cannot be the corresponding quasi-particle coordinate. Instead in
the next section we
search for an appropriate non-linear combination of ${\bf A}_L$ and $\rho$.

\section{Search for the longitudinal quasi-particle}

We search for nonlinear combinations of
the longitudinal vector potential and $\rho$ for which in
(\ref{nonfactlong}) and (\ref{nonGauss}) equalities can be verified. 
This should support
the general view of almost free quasi-particle composition of 
finite temperature media.  It should be checked whether 
for this new combination the equipartition and the spectral equation
of state can be interpreted consistently under the assumption of the
factorisation of their statistics. It
will be also checked to what extent it follows Gaussian statistics. 
Eventually we show
that the quality of the equation of state based on this independent
new variable is almost as good as the result of Section 2.
We shall work out the example of (\ref{nonlin}), and only comment on
the performance of the alternative proposition (\ref{nonlin2}).

\begin{figure}
\begin{center}
\includegraphics[width=8.5cm]{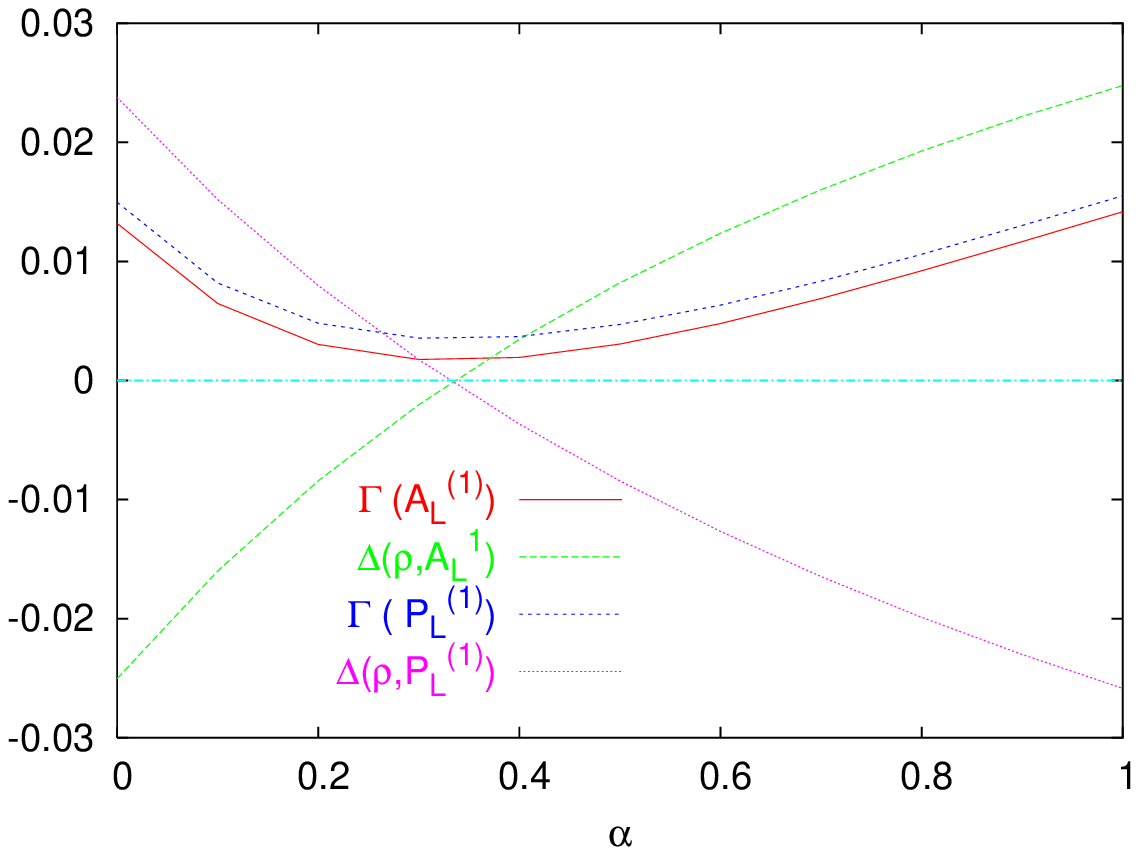} 
\includegraphics[width=8.5cm]{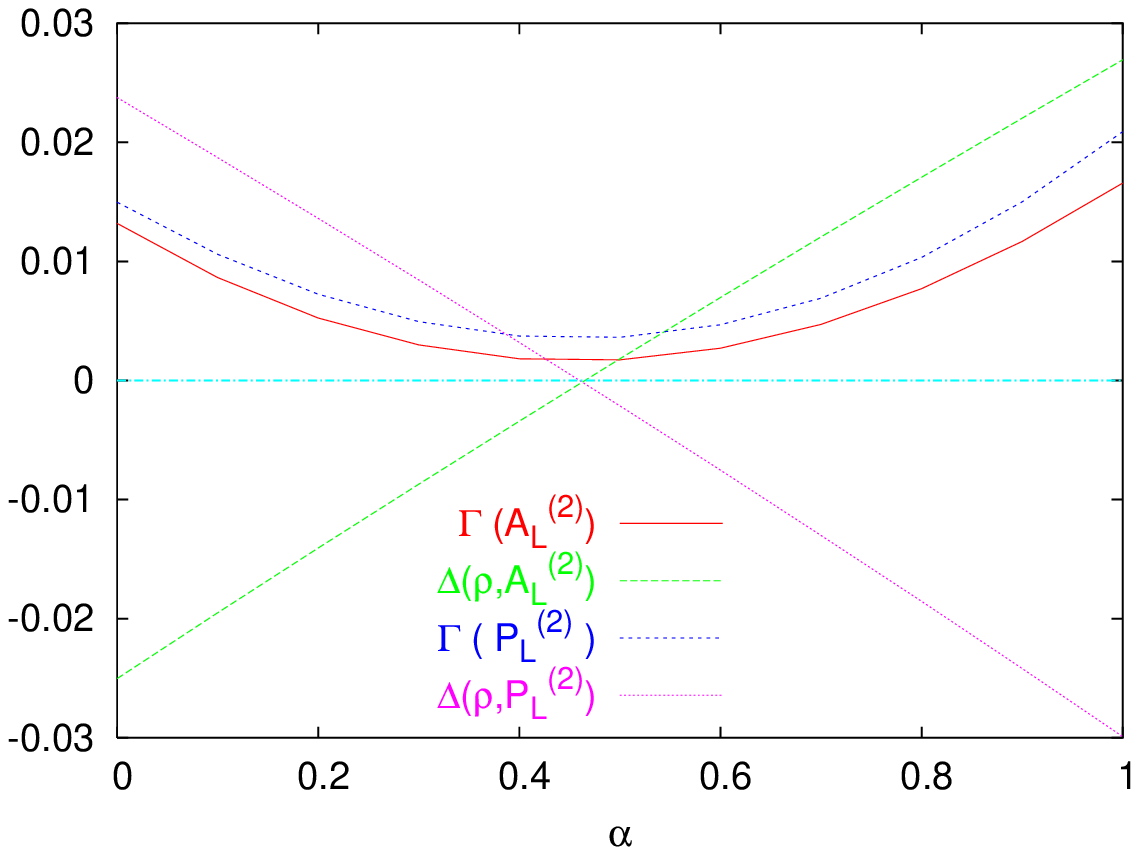}
\end{center}
\caption{Correlators of two versions of the nonlinearly mapped ''longitudinal''
field as a function of the parameter $\alpha$. On the left the
results of the mapping (\ref{nonlin}), on the right the mapping 
(\ref{nonlin2}) are shown.}
\label{kep:alfamer}
\end{figure}

The coefficient $\alpha$ in (\ref{nonlin}) could be estimated
analytically by assuming that it is small. 
For a quantitative treatment there is no need to apply
approximations. By tuning $\alpha$
continuously we have measured the normalised reduced correlation functions 
$\Delta({\cal A}_L,\rho)$ and $\Delta({\cal P}_L,\rho)$. 
In Fig. \ref{kep:alfamer} these two curves
are displayed as functions of $\alpha$ together with the quantities
testing the Gaussian nature of the new variables. It is remarkable
that both the new coordinate and its conjugate momentum
become uncorrelated with the Higgs-field for the same
$\alpha$. Although they are not perfect Gaussian variables, the deviation from
Gaussianity is minimal just for this $\alpha$ value, it is about 5 times
smaller than at $\alpha=0$. The situation is
very similar for the nonlinear mapping (\ref{nonlin2}) as can be seen
from the twin figure of Fig. \ref{kep:alfamer}. It is important to
notice that in the
latter case $\alpha\neq 1$, which would be the naive expectation on
the basis of the way the kinetic-potential equipartition is fulfilled.
Further in the analysis we use the optimal $\alpha$ values found as
the average over 40 independent configurations. It has to
be emphasised that no unique $\alpha_{opt}$ exists for single
configurations, the optimal values found from the four different
criteria fluctuate somewhat. It is the ensemble average which leads to
the remarkable coincidence displayed in Fig. \ref{kep:alfamer}.

\begin{figure}
\begin{center}
\includegraphics[width=12cm]{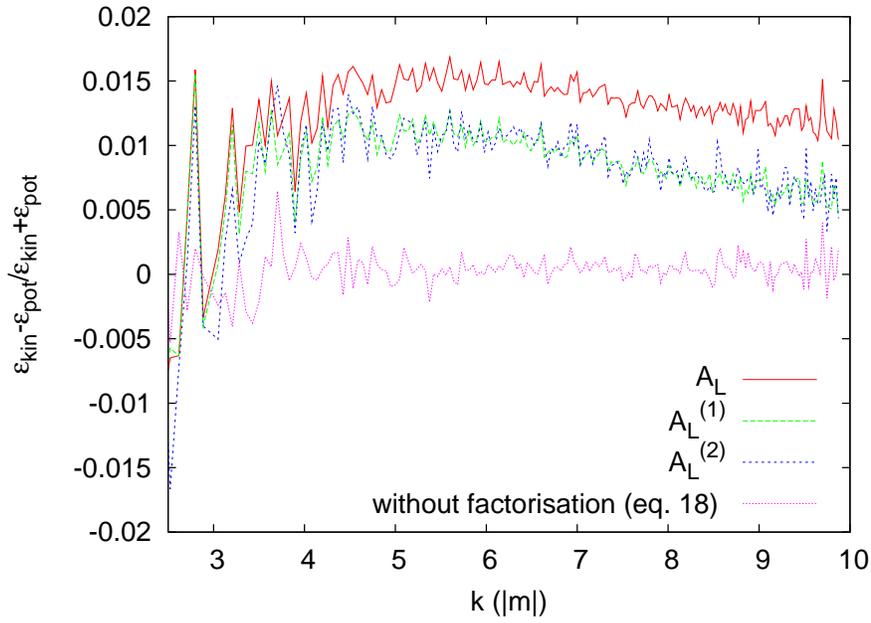} 
\end{center}
\caption{ The quality of the mode-by-mode equality of the
  longitudinal kinetic and potential energies 
when i) the original ${\bf A}_L$, ii) 
the nonlinearly constructed variable ${\cal
  A}_L^{(i)}$ are proposed as statistically independent fields.
For comparison the  perfect fullfillment of (\ref{equilong}) is also
  presented . 
The region $k\geq 2|m|$ is shown, where the equipartition is
very well fullfilled for the other degrees of freedom. }
\label{kep:ekvipart}
\end{figure}

Now one has to reexpress the partial energy densities and pressures in
terms of the new variables.
One should not forget that as a consequence of (\ref{nonlin}) also the
canonical momentum conjugate to $\rho$ changes:
\be
{\cal P}_\rho+\frac{2\alpha\rho}{1+\alpha\rho^2}{\cal A}_L{\cal
  P}_L={\bf \Pi}_\rho
\ee
Assuming factorisation of the statistical averaging over $\rho, \Pi_\rho$ and
${\cal A}_L, {\cal P}_L$ one obtains the following expressions for the averaged
energy densities:
\bea
\overline{\epsilon}_\rho&=&\frac{1}{2}\overline{{\cal P}_\rho^2}+
\overline{\frac{2\alpha^2\rho^2}{3(1+\alpha\rho^2)^2}}\overline{{\cal
    A}_L^2}~\overline{{\cal
    P}_L^2}+\frac{1}{2}\overline{(\nabla\rho)^2}+\frac{1}{2}
M_{eff}^2\overline{\rho^2},\nonumber\\
\overline{\epsilon}_L&=&\frac{1}{2}\left(\overline{(1+\alpha\rho^2)^2}+
\overline{\frac{(\alpha\nabla\rho^2)^2}{e^2\rho^2}}\right)
\overline{{\cal P}_L^2}+\frac{1}{2}
\overline{\frac{(1+\alpha\rho^2)^2}{e^2\rho^2}}\overline{(\nabla{\cal
    P}_L)^2}\nonumber\\
&+&\frac{1}{2}\overline{\frac{e^2\rho^2}{(1+\alpha\rho^2)^2}}
\overline{{\cal A}_L^2}.
\label{energynonlin}
\eea
It is worth now to check if the equality between the first two and
the last term of Eq.(\ref{energynonlin}) is fullfilled in
the present factorised form. Fig. \ref{kep:ekvipart} demonstrates no
more than 1\% deviation from the almost perfect equality found in
Section 2. 

The corresponding expressions for the partial pressures are as follows:
\bea
\overline{p}_\rho&=&\frac{1}{2}\overline{{\cal P}_\rho^2}+
\overline{\frac{2\alpha^2\rho^2}{3(1+\alpha\rho^2)^2}}\overline{{\cal
    A}_L^2}~\overline{{\cal
    P}_L^2}-\frac{1}{6}\overline{(\nabla\rho)^2}-\frac{1}{2}
M_{eff}^2\overline{\rho^2},\nonumber\\
\overline{p}_L&=&\frac{1}{6}\left(\overline{(1+\alpha\rho^2)^2}~\overline{{\cal
  P}_L^2}-\overline{\frac{e^2\rho^2}{(1+\alpha\rho^2)^2}}
\overline{{\cal A}_L^2}\right)\nonumber\\
&+&\frac{1}{2}
\overline{\frac{(\alpha\nabla\rho^2)^2}{e^2\rho^2}}\overline{{\cal
    P}_L^2}+\frac{1}{2}\overline{\frac{(1+\alpha\rho^2)^2}{e^2\rho^2}}
\overline{(\nabla{\cal P}_L)^2}.
\label{pressurenonlin}
\eea
It is worth to write explicitly the spectral equation of state
resulting for the new effective field variables, when the improved but
still approximate equipartition is exploited:
\be
\displaystyle
 w_{new}(k)=\frac{1}{3}\frac{\overline{\frac{(\alpha\nabla\rho^2)^2}
{e^2\rho^2}}+{\bf k}^2\overline{\frac{(1+\alpha\rho^2)^2}{e^2\rho^2}}}
{\overline{(1+\alpha\rho^2)^2}+\overline{\frac{(\alpha\nabla\rho^2)^2}
{e^2\rho^2}}+{\bf k}^2\overline{\frac{(1+\alpha\rho^2)^2}{e^2\rho^2}}}.
\label{eqstatnonlintheory}
\ee

\begin{figure}
\begin{center}
\includegraphics[width=12cm]{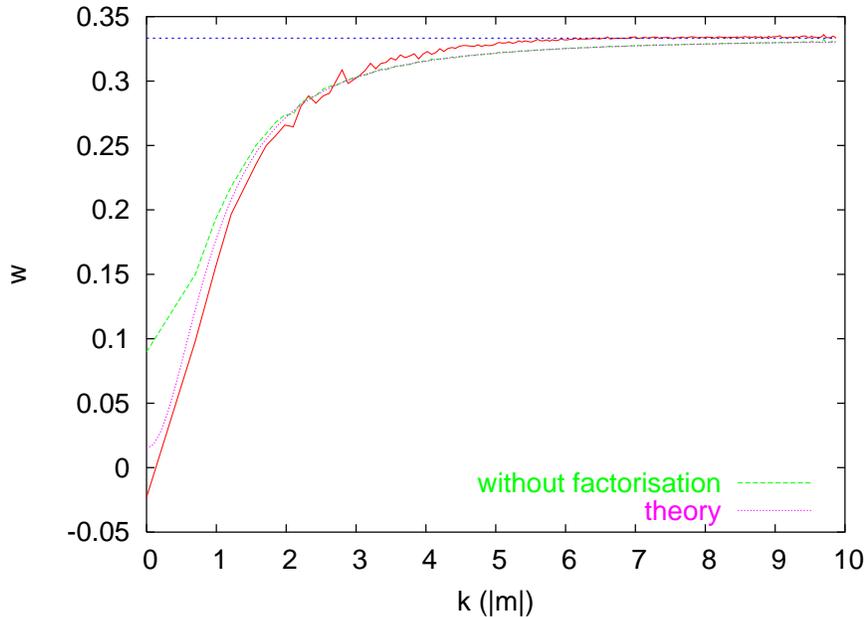} 
\end{center}
\caption{ Sperctral equation of state with ${\bf {\cal A}}_L^{(1)}$ 
(continuous line) as
  compared with the curve (\ref{eqstatnonlintheory}) (''theory'', dotted line)
  where the $\rho$ dependence of the coefficients is averaged numerically.
 The equation of state obtained in Section 2 (''without
 factorisation'', dashed line)
  is also shown for comparison.}
\label{kep:theory}
\end{figure}

The correction in the energy density and the pressure of $\rho$ turns
out to be negligible.
In Fig. \ref{kep:theory} we plot the ''longitudinal'' equation of state using
(\ref{energynonlin}) and (\ref{pressurenonlin}), by measuring the
corresponding averages directly, and the ''theoretical'' curve
(\ref{eqstatnonlintheory}). A rather remarkable agreement
of the curves is seen which works the best for
 low and moderate values of $k$. The systematic deviation
experienced in the high $k$ region would require smaller mass, which
however would lead to large discrepancy in the low $k$ region.
Similar features are observed when the equation of state is
constructed with help of the mapping (\ref{nonlin2}). The curve
corresponding to (\ref{eqstatnonlintheory}) agrees very well with the
longitudinal equation of state appearing in Fig. \ref{kep:wekvi}.

\section{Conclusions}
In the present investigation we searched for the quasi-particles of
the classical Abelian Higgs model at finite (low) energy density. The
validity of a quasi-particle representation of the thermodynamics
 is essential for cosmological
applications.  The intuitively expected statistical independence and
Gaussian behaviour of the transversal gauge field
and of the Higgs excitations in the unitary gauge was confirmed.

The longitudinal component of the vector field is not independent
statistically at finite energy density from the Higgs oscillations. This
correlation and the deviation from the Gaussian statistics grows with
the temperature and also with the $\lambda/e^2$ ratio. 
Still, its thermodynamical characteristics represented in form of a
spectral equation of state were shown to be
degenerate with the features of the transverse vector modes.

The first step of the search for an independent Gaussian pair of
canonically conjugate field variables describing the non-transversal
gauge fluctuations was put forward in this paper. We are confident
that a fourth quasi-particle degree of freedom completes the
thermodynamical characterisation, though there is no unique
prescription for its systematical construction. One possibility is to
abandon the idea of a local mapping in ${\bf x}$-space and try to
use a ${\bf k}$-dependent mixing parameter: $\alpha({\bf k})$.

The present experience is very useful also for the
investigation of the real time evolution of the field excitations in
 an inflaton+Gauge+Higgs system
starting from an unstable (symmetric) initial state and ending in the
broken symmetry phase. We are interested in the rate of excitation of
the different field degrees of freedom immediately after the exit from
inflation. Already at present it is clear that
vortex-antivortex pair production will be rather frequent and can be
studied also with help of the correlation coefficients introduced in
the present study. It will be interesting to see whether the energy
transfer from the inflaton to the quasi-particles related to the
transverse and the longitudinal vector fields is symmetrical.
 These questions are under
current investigation.

\section*{Appendix}

The thermal equilibrium of the system was realised in our real time
simulations by long temporal
evolution. It started from a noisy initial state having finite energy
density. In order to implement the equations of motion in the temporal
axial gauge, a spatial lattice was introduced. On this lattice the
gauge field and the covariant (forward, $F$) 
derivatives of the scalar field are approximated with help of link variables:
\be
U_i(x)=e^{-ie|a_i|A_i(x)},\quad D_i^F\Phi (x)=\frac{1}{|a_i|}\left(U_i(x)\Phi
(x+a_i)- \Phi (x)\right).
\ee
The compact nature of $U$'s has no effect at very low energy densities
studied in the persent investigation.
The solution of the equations of motion was started with all conjugate
momenta set to zero.
Under this condition the Gauss constraint is trivially
fulfilled in the temporal axial gauge and it is preserved during the
evolution. 
The Fourier modes of the canonical coordinate fields were given an amplitude
with random phase, providing for each mode equal potential energy.
The simulations were done mostly on $N=32$ three-dimensional
lattices with a lattice constant $a=0.35|m|^{-1}$. In order to test the
possible lattice constant dependence of the effects the system was
solved also on $N=64, a=0.174|m|^{-1}$ lattice. Notable lattice
spacing dependence was observed only for the Higgs mass (see the main
discussion).

The vector potential and its conjugate momenta are computed on an
isotropic lattice $(a_i=a)$ as
\be
A_i(x)=-\frac{1}{ea}{\rm Im}U_i(x),\quad \Pi_i=\frac{1}{ea\delta
  t}{\rm Im}\left(U_i(x+a\delta t)U^{-1}_i(x)\right),
\ee
where the discrete time-step we use is $a_0=a\delta t$ with $\delta
t\sim 0.1$. These fields were decomposed into tranverse and
longitudinal components, we need in the formulae for the
decomposition of the energy and pressure.
The magnetic field strength is measured from the plaquette variables
\bea
U_{ij}&=&U_j^{-1}(x)U_i^{-1}(x+a_j)U_j(x+a_i)U_i(x),\nonumber\\
{\rm Im}U_{ij}&=&-ea^2\epsilon_{ijk}B_k.
\eea

 In the moments of measurements the actual configuration was
transformed to unitary gauge.

\section*{Acknowledgements}
The authors are grateful to J. Smit for valuable suggestions.  
Enjoyable discussions with Sz. Bors\'anyi, A. Jakov\'ac and P. Petreczky
are also acknowledged. This research was supported by the Hungarian
Research Fund (OTKA), contract No. T-037689.

\end{document}

%% file: masstabl2.tab
0.0409 & 1.91 & $ 1.80 \pm 0.05 $ & $ 0.93 \pm 0.02 $ & $  2.16 \pm 0.02 $ & $ 0.95 \pm 0.01 $ & $ 0.92 \pm 0.01 $ \\ 
0.0102 & 0.47 & $ 1.95 \pm 0.03 $ & $ 0.98 \pm 0.01 $ & $  2.23 \pm 0.02 $ & $ 0.99 \pm 0.02 $ & $ 0.98 \pm 0.01 $ \\ 
0.0025 & 0.12 & $ 1.99 \pm 0.01 $ & $ 1.00 \pm 0.00 $ & $  2.26 \pm 0.03 $ & $ 1.00 \pm 0.02 $ & $ 0.99 \pm 0.01 $ \\ 
0.0006 & 0.03 & $ 2.00 \pm 0.00 $ & $ 1.00 \pm 0.00 $ & $  2.26 \pm 0.03 $ & $ 1.00 \pm 0.03 $ & $ 1.00 \pm 0.01 $ \\ 